# Energy-Efficient Tethered UAV Deployment in B5G for Smart Environments and Disaster Recovery

Abdu Saif[1], Kaharudin Dimyati[1], Kamarul Ariffin Noordin[1], Nor Shahida Mohd Shah[2], S. H. Alsamhi[2] and Qazwan Abdullah[3]

[1]Department of Electrical Engineering University of Malaya  Kuala Lumpur, Malaysia
[2] Faculty of Engineering Technology Universiti Tun Hussein Onn Malaysia Johor, Malaysia
[3] Software Research InstituteAthlone Institute of Technology  Athlone, Ireland & IBB University, Ibb, Yemen

*Abstract*— Due to Unmanned aerial vehicles (UAVs) limitations in processing power and battery lifetime. The tethered UAV (TUAV) offers an attractive approach to answer these shortcomings. Since a tethered connected to UAV is one potential energy solution to provide a stable power supply that connects to the ground would achieve impressive performances in smart environments and disaster recovery. The proposed solution is intended to provide stable energy and increase the coverage area of TUAV for smart environments and disaster recovery. This paper proposed that the tethered connected to UAV will provide the continuous supply and exchange the data with ground terminals.  Besides the adjustable tether length, elevation angels act to increase the hovering region, leading to the scalability of coverage in many applications.

Moreover, the power consumption and transmission distance while achieving a trade-off between the hovering and coverage probabilities. The simulation results demonstrate efficient performance in terms of line-of-sight probability, path loss, and coverage probability for scalability coverage smart environments and disaster recovery scenarios. Furthermore, maximum coverage probability is achieved versus increased tethered length because of the gain and fly over a region of maximum tethered.

## I. Introduction

Natural disasters such as earthquakes, hurricanes, tornadoes, and severe snowstorms frequently result in devastating telecommunication infrastructures[1]. Increasing the mortality rate during rescue from natural disasters is a significant challenge [2],[3]. Therefore, growing future demands further boost wireless network communication in a large area rapidly under infrastructure failure [4]. There is a robust demand for public safety communications between first responders and victims for search and rescue operations [5]. However, flexibility, low-latency services, and swift adaptation to the environment public safety are required to provide wireless coverage services during natural disasters [6]. Therefore, replacing the Ground Base Station (GBS) with an Unmanned Aerial Vehicle (UAV) is viable and integrated with optimal relay hops to improve wireless coverage services [7]. The Beyond fifth Generation (B5G) technology evolution features help in blockchain-based UAV networks such as coverage Extension, Reliable connectivity, and low-power consumption [8]. UAVs can be used as mobile base stations to provide wireless coverage services in smart environments [9] and disaster-stricken areas [6]. However, the primary drawback is that UAVs run on battery power which can run out very quickly [10].

Consequently, UAVs' energy consumption and battery life become a significant constraint in the case of network infrastructure collapse [11]. Moreover, UAV flying is a limited battery capacity during coverage services in disaster scenarios [12]. Furthermore, gathering data from the Internet of Things (IoT) using UAVs also suffers from data processing and transferring due to limited battery charge [13],[14]. Therefore, the authors of [15] applied an artificial neural network for predicting the signal strength over IoT devices in smart environments based on optimized location and UAV trajectory.  Furthermore, UAV is used for green smart environments in B5G networks [9]. Natural disasters could lead to significant service disruptions since the B5G mission does not focus on post-disaster emergency communication scenarios [16],[17]. Tethered balloon plays a vital role for improving green communication in smart cities [18], extending coverage area [19]  and deliver broadband communication services in disaster areas [20].

UAV is considered a relay station [9]. Tethered represents the critical solution for providing power supply to UAVs. Tethered is used to tie UAV to the ground, fast data transfer, supply power to UAV, and solve the battery lifetime   [21]. The UAVs coverage area defines the locations of active ground user devices that get the coverage services from the UAVs. Moreover, tethered can enhance flight path gap areas

Table.1 Comparison of existing works

| Ref. | Contribution | 1 | 2 | 3 | 4 | 5 |
|---|---|---|---|---|---|---|
| [21](2019) | Tethered balloon for an efficient emergency communication system, reducing casualty mortality and morbidity for disaster recovery. | X | √ | √ | X | √ |
| [22](2019) | Tethered balloon technology is utilized for delivering broadband services in a disaster event. | X | √ | √ | X | √ |
| [23](2020) | The optimal placement of TUAVs is investigated to minimize the average path loss to the receiver located on the ground. | √ | √ | X | √ | √ |
| [24](2020) | utilization of the TUAV to assist the cellular network. | √ | √ | X | X | √ |
| [25](2019) | UAV-enabled cellular network setup based on tethered TUAVs | √ | X | X | √ | √ |
| Our work | We are focused on TUAVs and the effect of UAV hovering to improve TUAVs coverage area in smart environments and disaster areas. | √ | √ | √ | √ | √ |

1- UAV  2- Tethered 3- ECS   4- Tethered elevation angle   5- Coverage improvement

and increase the covered regions [22]. Therefore, to improves the hovering region's UAV flight path and maximizes the coverage area, tethered UAV (TUAV) is needed [23].

UAV is connected to tethered, which can fly for a long time according to the purposes and support heavier payload [24]. Also, UAVs can send the gathered data via a tethered data link to enable backhaul connectivity for carrying power and data to reliable connectivity [26],[27]. Due to the presence of a wired data connection via the tether, the authors in [28] proposed used UAVs connected to tethered in post-disaster scenarios to provide the backhaul connectivity for the set of UAVs. In [29], the optimal location of UAV helps to maximize the number of covered per user devices with lower path loss. The authors of [30] addressed path-loss module behaviors from a cellular base station toward a flying UAV. The path-loss value is a function of tether node, user devices' elevation angles, and distances from the UAV [8]. Due to UAV's mobility freedom, it faces limited battery and flies time [31]. The selected hovering UAV based on the angle of the tether node was discussed in [25].

A. Contributions

This paper proposed a system model for the TUAV-assisted smart environments and disaster areas in B5G networks that enable energy efficiency. The critical aspect of our approach is mainly focused on TUAV angles to deliver communication service to extensive coverage and longtime. First, the proposed TUAV for smart environments and disaster areas evaluate based on analyzing the impacts of tethered angle to TUAV hovering angels, the path loss, and coverage probability. We then investigated the trade-off between the hovering and coverage regions based on the tethered length to improve coverage area. Finally, we analyzed the LoS, path loss, and coverage probability versus the tethered elevation angle for sustainable operations.

B. Paper structure

The remainder of the paper is organized as follows. Section II presents the system model. Section III offers the simulation of results for further analysis of the UAV. Finally, section IV concludes the paper.

II. SYSTEM MODEL

This section presents the system model used to implement the TUAV approach for coverage improvement. First, a set of user devices in smart environments or in disaster areas unable to obtain wireless coverage from the terrestrial communication network is considered. Then, TUAV is deployed to provide wireless coverage service to the user device. TUAV supports smart environments and a disaster recovery for longtime and significant coverage. Therefore, TUAV is connected to tether an optical prototype ground unit while preserving its energy consumption, as shown in Fig. 1. Finally, the hovering rejoins define as the area of TUAVs' trajectory path to provide the coverage service for ground user devices. The tethered has the freedom to move in the hovering region to maximize the coverage region based on tether length ($T_{max}$). The cartesian coordinates for TUAV nodes and user devices are represented as $x_j, y_j, z_j$), $(x_k, y_k)$, respectively. The tether nodes are connected to the optical fiber in the ground that located at $(x_i, y_i)$. Hence, tethered provides a power supply to UAVs and processes data gathered from disaster and smart environments in real-time. In addition, the hovering region will be achievable through tethered in 3D to improve the large user device scale coverage areas. In particular, when the distance between the user devices located in smart environments or disaster areas is so much, the tethered length will be adjustable to a certain threshold to provide efficient coverage services to user devices.

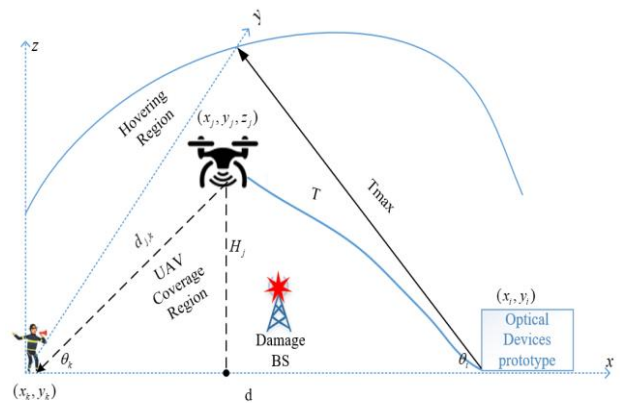

Fig. 1. System Model

The reachable hovering region of the TUAV is connected to tethered in 3D, denoted as:

$$R = \begin{cases} (x_j - d)^2 + y_j^2 + z_j^2 \leq T_{max} \\ sin^{-1}\left(\frac{z_j^2}{(x_j-d)^2+y_j^2+z_j^2}\right) \geq \theta_i \end{cases} \quad (1)$$

Where Tmax is the maximum of tethered length, $\theta_i$: tethered elevation angle. Hence, the tethered is aimed to obtain an optimal location within the hovering region (R) to the maximum coverage probability.

## A. Line of sight for TUAV and User devices

The LoS probability of TUAV connected to tethered affect the user device receivers as follows [23]:

$$P_{LoS} = a\left(tan^{-1}\left(\frac{z_j}{\sqrt{x_j^2+y_j^2}}\right)\right)^b \quad (2)$$

The possibility of an LoS and an NLoS for the k[th] user devices is a function of the user's elevation angles and TUAV altitudes. Then the TUAV to user devices communication link is shown as a function of $P_{LoS}(\theta_k)$ and $P_{NLoS}(\theta_k)$ as follows [32]:

$$P_{LoS}(k) = \frac{1}{1+a\,exp(-b(\theta_k,h)-a)} \quad (3)$$

$$P_{NLoS}(k) = 1 - P_{LoS}(k) \quad (4)$$

where $\theta_k$ is the elevation angles of user devices in the disaster area. $a$ and $b$ are parameters that affect the S-curve parameters and are varied according to the smart environments, such as urban, suburban, dense urban, and high-rise urban.

## B. Analysis of Path Loss (PL)

Path loss propagation is a critical factor that affects the wireless channel, including the attenuation of radiated signals with distance, user device elevation angle, and the distance from the UAV to user devices. Therefore, the pathless of user devices include LoS and NLoS links with multipath fading and shadowing from the transmitted signals due to blocking and large-scale path loss obstacles. From Eq (3), the communication link is more likely to be an LoS communication link with more significant user devices elevation angles. Subsequently, the average PL between a UAV and the user devices are denoted as follows [12, 33],[34]:

$$PL(dB) = \left(\frac{4\pi f_c d_{j,k}}{c}\right)^2 (P_{LoS}\eta_{LoS} + P_{NLoS}\eta_{NLoS}) \quad (5)$$

where d: The distance between the UAV and user devices, ($\eta_{LoS}$, $\eta_{NLoS}$) are additional PL values for LoS and Non-LoS. $f_c$ denotes as carrier frequency, $c$ speed of light. Thus, the distance between the UAV and user devices nodes indicated as:

$$d_{j,k} = \left((x_j-x_j)^2 + (y_j-y_i)^2 + z_j^2\right)^{\frac{1}{2}} \quad (6)$$

## C. Coverage Probability of Downlink

The achievable coverage probability of TUAV to user devices in disaster areas is given by Eq (7). When the user devices are located in the coverage region, the condition coverage probability achieved by the user devices distance can be expressed as follows:

$$P_{cov} = \begin{cases} P_{LoS}Q(\frac{p_{min}+PL(dB)-p_t-GdB+\mu_{LoS}}{\sigma^2_{LoS}}) + \\ P_{NLoS}Q(\frac{p_{min}+PL(dB)-p_t-GdB+\mu_{NLoS}}{\sigma^2_{NLoS}}) \end{cases} \quad (7)$$

Where PL(dB) represents the path loss, $\sigma^2$ is noise power, $GdB = 3dB$ represents the antenna gain, and $Q$ is the function, and $\mu$ is additional path loss for Los and NLoS links.

TABLE I. SIMULATION PARAMETERS

| Parameters | a | b | $\mu_{LoS}$ | $\mu_{NLoS}$ |
|---|---|---|---|---|
| Urban | 10.6 | 0.18 | 1 | 20 |
| Tethered Elevation Angles | 20º, 30º, 60º | | | |
| Noise power ($\sigma^2$) | -174 [dBm] | | | |
| UAV transmsstion power | 10 W | | | |
| Channel Bandwidth | 5MHz | | | |

## III. NUMERICAL RESULTS AND ANALYSIS

In this section, the simulation results have been presented to demonstrate the performance of the proposed method. The tethered power supply will be analyzed for several elevation angles of tethered linked with TUAV at the ground nodes. Table I shows parameters used in the proposed method to select the optimal TUAV placement for the varied distance between TUAV and user devices nodes. Subsequently, probability LoS/NLoS is usually modelled to deliver wireless signals from TUAV to user devices in the downlink. In addition, the performance of the path loss and coverage probability versus tethered elevation angels are considered based on the tethered length and distances from TUAV to ground terminals.

Fig 2. shows that the performance of LoS probability increased when the distance between tethered nodes connected to TUAV and user devices simultaneously increased for the same level of hovering and coverage regions across the elevation angles. It can be seen the maximum probability of LoS achieves at a distance of 500 m in the case of (Tethered angle ($\theta_i = 20º$). Due to the altitude of TUAV adjusted by the length of tethered. Thus, tethered length can increase the gain and fly over a region and operate optimally within the receiver's LoS range. Therefore, when (Tethered angle ($\theta_i$)) Was increased to 30º and 60º, respectively. The LoS reach the maximum value at a distance (600, 900) m. This is a fact due to the increased $\theta_i$ impacts the SNR of user devices through increased UAV altitudes and the user devices and increased the UAV coverage area.

Conversely, Fig 3. shows that the performance of NLoS probability decreased when the tethered nodes to user devices distance increased across different $\theta_i$. The minimum probability of NLoS was achieved at a distance of 550 m with (Tethered angle ($\theta_i = 20º$), and distance 600 m with (Tethered angle ($\theta_i = 30º$). Due to the loss of obstacles, it receives signals at the user devices nodes. Moreover, the minimize achievable LoS probability at a distance of 900m when tethered elevation angle increases to $\theta_i = 60º$. This is due to the large scale of NLoS, path loss that affected user devices SNR.

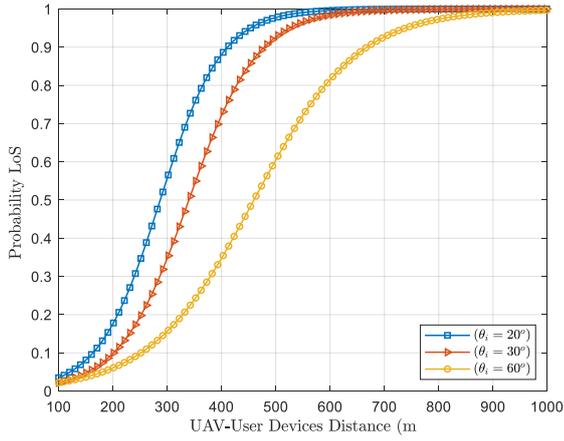

Fig. 2. Probability LoS versus distance cross with different (Tethered angle ($\theta_i$)).

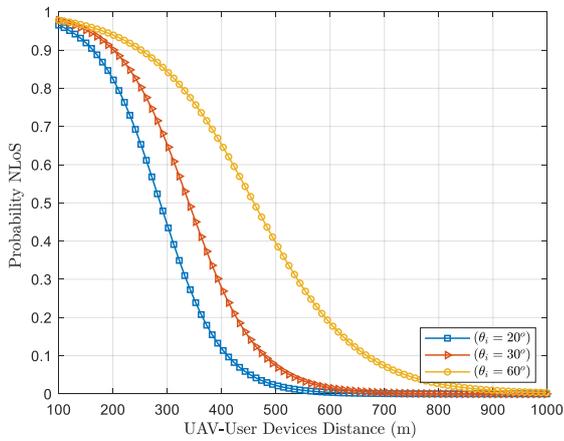

Fig. 3. Probability NLoS versus distance cross with different (Tethered angle ($\theta_i$)).

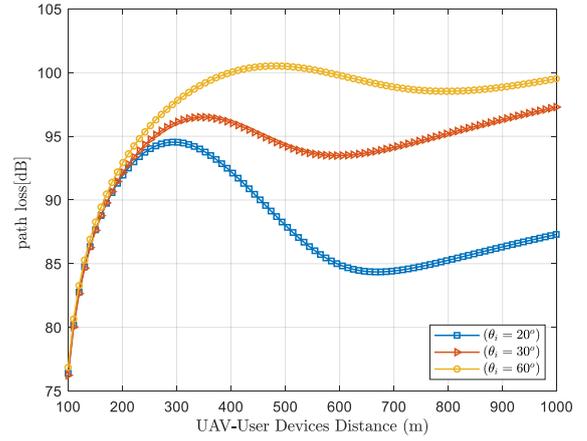

Fig. 4. Path loss versus distance cross with different (Tethered angle ($\theta_i$)).

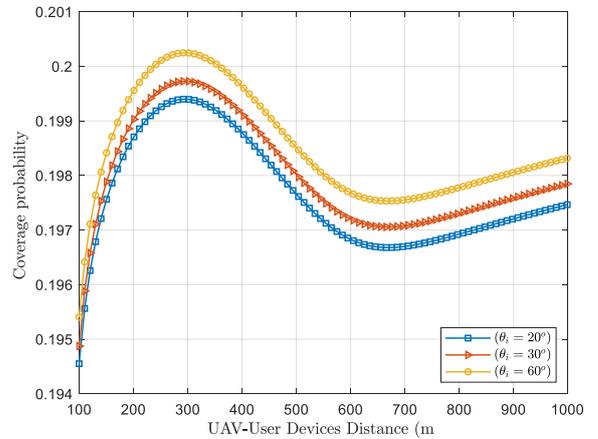

Fig.5 Coverage probability versus distance at different (Tethered angle ($\theta_i$)).

Fig 4. shows that path loss was affected by the distance between the tethered node in TUAV and user devices into (Tethered angle ($\theta_i$)). The path loss was maximized from 75 dB to 105 dB, and the varied distance increased from 100 m to 500 m at (Tethered angle ($\theta_i = 20°$)), due to the power consumption of user devices. Therefore, the path loss decreased from maximum value to 96 dB and 94 dB when the distances of tethered nodes to user devices increased from 100 m to 300 m for (Tethered angle ($\theta_i = 30°, 60°$)) respectively. Finally, a lower path loss and small-scale fading have been found in the ATG channel while having a longer link length that deteriorates the received SNR.

Fig 5. shows the performance of the coverage probability versus distances between the tethered nodes to user devices distance cress with (Tethered angle ($\theta_i$)). It can be seen that maximum coverage probability is achieved at 300m due to the increasing LoS and NLoS coverage at user device nodes. However, lower coverage probability was found at a distance of 650m in the case of all varied (Tethered angle ($\theta_i$)). This is due to received single only from LoS at the destination nodes and slow LoS for those changes in (Tethered angle ($\theta_i$)). Therefore, the coverage probability increased at a distance of more than 700m for those distances is still active TUAV coverage areas.

IV. Conclusion

This paper discussed the TUAV power supply and gathered data to improve smart environments and disaster coverage and delivering communication service for a long time. The tethered helps UAV for visibility links between in/out-coverage areas. TUAV plays a vital role in many applications in smart environments such as smart cities, smart homes, smart agriculture, smart healthcare, smart manufacturing, disaster recovery. Then, The approach has been implemented as an adjustable of (Tethered $\theta_i$) to improve coverage probability from TUAV over smart environments and disaster areas. The proposed model provides a good performance LoS and reduces path loss of the system. Besides, the obtained results in term of TUAV coverage probability was improved significantly.

ACKNOWLEDGEMENT

This work is supported by the University of Malaya, Department of Electrical Engineering, and funded by the DARE project (Grand ID: IF035A-2017 & IF035-2017).